\documentstyle[aps,pra,twocolumn]{revtex}
\begin{document}
\draft

\title{
Structure of nonlinear gauge transformations
}
\author{Marek Czachor}
\address{
Katedra Fizyki Teoretycznej i Metod Matematycznych\\
 Politechnika Gda\'{n}ska,
ul. Narutowicza 11/12, 80-952 Gda\'{n}sk, Poland
}
\maketitle
\begin{abstract}
Nonlinear Doebner-Goldin [Phys. Rev. A {\bf 54}, 3764 (1996)] 
gauge transformations (NGT) defined in terms of 
a wave function $\psi(x)$ do not form a group. To get a group
property one has 
to consider transformations that act differently on different
branches of the complex argument function 
and the knowledge of the value of $\psi(x)$
is not sufficient for a well defined NGT. NGT that are well
defined in terms of $\psi(x)$ form a semigroup parametrized by
a real number $\gamma$ and a nonzero $\lambda$ which is either an
integer or $-1\leq \lambda\leq 1$. An extension of NGT to
projectors and general density matrices leads to NGT with
complex $\gamma$. Both linearity of evolution
and Hermiticity of density matrices are gauge dependent properties.
\end{abstract}

Adopting the view that {\it all\/} actual measurements of quantum
mechanical systems are eventually reducible to those of
particle positions at various moments of time
\cite{F,M,dE,W,R} one has to accept
a new gauge principle: All theories that give the same
probability densities in {\it position\/} space for all times
and all experimental arrangements are indistinguishable. 
Once we agree on this viewpoint the first question we have to answer is
what is the most general set of transformations that leave
probability densities in position space invariant. Certainly
this set is bigger than just the unitary and anti-unitary
ones. A class of {\it nonlinear\/} transformations of {\it wave
functions\/} that have this property was extensively
investigated by Doebner and Goldin, and
their collaborators from the Clausthal school 
\cite{DGpra,DG1,DG2,DG3,DG4,DG5,DG6,DG7,DG8}. 
The Doebner-Goldin transformations are sometimes referred to as
gauge transformations of the third kind.

Taking a linear Schr\"odinger equation and performing a gauge
transformation of the third kind one arrives at an equivalent
theory whose dynamical equation is in general nonlinear.
Linearity is therefore a gauge dependent feature and, as such,
cannot be physically essential. It follows that there exist nonlinear
theories which are physically indistinguishable from
standard quantum mechanics unless one finally invents an
experiment which is not reducible to a measurement of position. 

Gauge transformations of the third kind are defined at the level
of {\it wave functions\/} \cite{spin}  and are characterized by {\it two
real\/} parameters:
\begin{eqnarray}
{}&{}&\psi(x)\mapsto\nonumber\\
&{}& N_{\lambda,\gamma}[\psi](x)
=
|\psi(x)|
\exp
\big[
i\lambda\, {\rm arg\,} \psi(x)
+
i\gamma\ln |\psi(x)|
\big]. \label{gg}
\end{eqnarray}
Here arg denotes a phase of $\psi(x)$. This phase possesses a
$2n\pi$ ambiguity which is regarded as irrelevant \cite{DGpra}.
Complex conjugation is a particular case of (1) and is given by $N_{-1,0}$. 
The transformation (\ref{gg}) leaves the positional probability
density invariant 
\begin{eqnarray}
|N_{\lambda,\gamma}[\psi](x)|=|\psi(x)|.
\end{eqnarray}
If $\psi$ satisfies an ordinary linear
Schr\"odinger equation the transformed wave function is a
solution of some nonlinear Doebner-Goldin equation (compare
Eq.~(11); for a detailed discussion cf. \cite{DGpra}). Starting with a
solution $\psi$ of a Doebner-Goldin equation one arrives at an
equation belonging to the same class but with some parameters
suitably transformed. All equations obtained from each other by
$N_{\lambda,\gamma}$, with $\lambda$ and $\gamma$ $t$- and
$x$-independent, are 1-homogeneous in
$\psi$ and therefore invariant under multiplication of
$\psi$ by constant factors. This shows that the phase ambiguity inherent
in the definition of arg$\,\psi$ does not lead to any ambiguity
of the nonlinear modification of the corresponding 
Schr\"odinger equation.

The objective of this note is twofold. First, it will be shown
that the phase ambiguity leads to a problem with group
properties of nonlinear gauge transformations. 
This leads to two classes of transformations. One is
characterized by nonzero integer $\lambda$ and the other by
nonzero real $\lambda$'s satisfying $-1\leq \lambda\leq 1$. In
both cases one obtains a semigroup structure. 
Second, we shall
see that there exists a natural extension of the Doebner-Goldin
transformations involving {\it three\/} real parameters, but in
order to find it one has to start with projector (or, more
generally, density matrix) representation of states. 

To see how the phase ambiguity affects the group property 
consider $N_{3/2,0}$ and a point $x$ satisfying ${\rm
arg}\,\psi(x)=\pi/4$ (now arg denotes both here and in
(\ref{gg}) the unique principal branch of
the argument function Arg i.e. $-\pi<{\rm arg\,}z\leq \pi$). 
One finds that 
\begin{eqnarray}
N_{3/2,0}\circ
N_{3/2,0}[\psi](x)=N_{9/4,0}[\psi](x) 
\end{eqnarray}
which agrees with the
affine group property mentioned in \cite{DGpra}. 
Now consider a point $x'$ satisfying ${\rm
arg}\,\psi(x')=3\pi/4$. The argument of the transformed function
is ${\rm arg\,}N_{3/2,0}[\psi](x')=-7\pi/8$. Aplying again
$N_{3/2,0}$ to $N_{3/2,0}[\psi]$ at $x'$ we find 
${\rm arg\,}N_{3/2,0}\circ N_{3/2,0}[\psi](x')=11\pi/16$.
On the other hand ${\arg\,}N_{9/4,0}[\psi](x')=-5\pi/16$ and
therefore 
\begin{eqnarray}
N_{3/2,0}\circ
N_{3/2,0}[\psi](x')\neq N_{9/4,0}[\psi](x'). 
\end{eqnarray}
It should be stressed that this property of
$N_{\lambda,\gamma}[\psi]$ holds in general for a set of points
of non-zero measure. 

A closer look at the origin of this problem shows that
the affine property 
\begin{eqnarray}
N_{\lambda',\gamma'}\circ
N_{\lambda,\gamma}=
N_{\lambda'\lambda,\lambda'\gamma + \gamma'}.\label{group}
\end{eqnarray}
does not hold for non-integer $\lambda$ ($|\lambda|>1$) if one fixes a
concrete branch of Arg in the definition (\ref{gg}). 
One can obtain (\ref{group}) while maintaining the principal
branch of Arg in (\ref{gg}) provided $-1\leq\lambda\leq 1$, but
then one gets a {\it semigroup\/} structure. Similarly,
restricting $\lambda$'s to integers one obtains a semigroup
satisfying (\ref{group}). 

To obtain a group property one has to modify (\ref{gg}) in order
to keep track of the branches of Arg. This can be done by
introducing an additional integer quantum number and defining 
\begin{eqnarray}
{}&{}&N_{\lambda,\gamma}[\psi]\big(x,n(\lambda,\gamma,m,x)\big)=
|\psi(x,m)|\nonumber\\
&{}&\times
\exp
\big[
i\lambda\, {\rm arg\,} \psi(x,m)
+
2m\lambda\pi i
+
i\gamma\ln |\psi(x,m)|
\big] \label{ggn}
\end{eqnarray}
where $n(\lambda,\gamma,m,x)={\rm E}\Big(
\frac{\lambda\, {\rm arg\,} \psi(x,m)}{2\pi}
+
m\lambda
+
\frac{\gamma\ln |\psi(x,m)|}{2\pi }\Big)$ and E$\,(r)$ denotes
an integer part of $r$. The transformations form a group but the
price one has to pay is in the presence of the additional
integer-valued function $n(\lambda,\gamma,m,x)$. 

Alternatively, one can {\it start\/} with two functions $A(x)$ and
$B(x)$, and the lower-triangular representation of the affine
group: 
\begin{eqnarray}
\left(
\begin{array}{c}
A\\
B
\end{array}
\right)
\mapsto
\left(
\begin{array}{cc}
1 & 0\\
\gamma &\lambda
\end{array}
\right)
\left(
\begin{array}{c}
A\\
B
\end{array}
\right)=:
\left(
\begin{array}{c}
A'\\
B'
\end{array}
\right).\label{AB}
\end{eqnarray}
The transformation $(A,B)\mapsto(A',B')$ induces a nonlinear
action of the affine group in the space of wave functions
\begin{eqnarray}
\psi=\exp[A+iB]\mapsto \exp[A'+iB']=:\psi'
\end{eqnarray}
which is equivalent to NGT. 
This is an alternative way of introducing NGT and such NGT {\it do\/} form a
group. However, it is essential that we {\it start\/} here with
$A$ and $B$, and {\it define\/} NGT at their level. 
If one wants
to do this at the level of $\psi$ (i.e use Eq.~(1)) 
one will necessarily face the
$2n\pi$ ambiguity in the definition of $B$ and 
the group property is lost.
This clearly shows that (1) and (7) are inequivalent. 
Let us note that working at the
level of $A$ and $B$ we control the branches of Arg~$\psi$ 
since $\lambda B$ is given exactly and not modulo $2\pi$. 
The problem with the group property therefore disappears if one
sticks to the so-called hydrodynamic version of the
Schr\"odinger equation \cite{BCK} which does not make use of
$\psi$ but works entirely at the level of the (Hamilton-Jacobi)
variables $A$ and $B$.  

If one does not
want to introduce these additional structures and unambiguously
compose transformations of the type (1), one has to
choose either $-1\leq\lambda\leq 1$ and use the principal
branch of Arg, or restrict $\lambda$'s to nonzero integers. In
what follows we shall assume that one of these possibilities has
been chosen. In both cases the transformations form a semigroup.

Consider now the problem of the number of parameters
characterizing a nonlinear gauge transformation. 
The projective nature of quantum mechanical state space makes it
more natural to work with $|\psi\rangle\langle\psi|$ than with
$|\psi\rangle$. The overall phase ambiguity is then
automatically removed from the description and an extension from
pure states to mixtures is natural. Let us begin with 
the pure state density matrix (projector)
$\rho(x,y)=\psi(x)\overline{\psi(y)}$. The nonlinear gauge
transformation of the wave function induces the following
transformation of the corresponding projector 
\begin{eqnarray}
{}&{}&N_{\lambda,\gamma}[\rho](x,y)\nonumber\\
&{}&=\rho(x,y)\exp
\Biggl[
(\lambda -1){\rm Ln} \frac{\rho(x,y)}{|\rho(x,y)|}
+
\frac{i\gamma}{2}\ln \frac{\rho(x,x)}{\rho(y,y)}
\Biggr].\label{ggdm}
\end{eqnarray}
The assumption of a privileged role of the
positional measurements leads to the requirement that the
diagonal elements of density matrices in position space should
be unchanged. This condtion is satisfied not only by (\ref{ggdm})
but also by a larger class of transformations involving a {\it
complex\/} parameter $\gamma_c$. The transformed density
matrices 
$N_{\lambda,\gamma_c}[\rho](x,y)$ are non-Hermitian but
\begin{eqnarray}
N_{\lambda,\gamma_c}[\rho](x,x)=\rho(x,x).
\end{eqnarray}
It is interesting that the additional (imaginary) parameter in
$N_{\lambda,\gamma_c}[\rho]$ does not show up if one starts with
$\psi$. This is so because the logical structure of derivation of NGT is
different for $|\psi\rangle$ than for $|\psi\rangle\langle\psi|$:
For $|\psi\rangle$ one first performs a nonlinear transformation
and then complex conjugates, finally requiring $|\psi'(x)|=|\psi(x)|$; for
$|\psi\rangle\langle\psi|$ one first complex conjugates
$|\psi\rangle$ in order to get $\langle\psi|$, and then performs
the nonlinear transformation. 
This is precisely the reason why the constraints imposed on
NGT by the consequences of complex 
conjugation are different for general $\rho$ (including pure
states!) than for state vectors or wave functions. 

It is evident that once we agree that linearity is a gauge dependent property
we have to accept the same status of Hermiticity of density matrices.
A gauge transformed density matrix satisfies an equation that is
nonlinear and dissipative but physically may be
indistinguishable from a linear and nondissipative theory. 

The fact that NGT establishes an equivalence between classes of
nonlinear Sch\"rodinger equations and ordinary linear
quantum mechanics automatically refutes all arguments for a
fundamental impossibility of nonlinear extensions of the
standard theory (see also \cite{DGpra}). Indeed, the transformations with
$\lambda=1$ and $\gamma\neq 0$ only add a nonlinear term while
maintaining the form of the kinetic and potential terms in the
Hamiltonian. To see this explicitly (cf. \cite{DG8}) assume that
$\psi$ is a solution of a linear Schr\"odinger equation,
$\psi'=N_{1,\gamma}[\psi]$,  and
$\gamma$ is in general time dependent. Then 
\begin{eqnarray}
i\hbar\partial_t\psi' &=&\Big(-\frac{\hbar^2}{2m}\Delta+V\Big)\psi'+
\frac{\hbar^2\gamma}{4m}\Big(iR_2+2R_1
-2R_4\Big)\psi'\nonumber\\ 
&\phantom =&
-
\frac{\hbar^2\gamma^2}{8m}\Big(2R_2-R_5\Big)\psi'
-
\frac{1}{2}\dot \gamma\ln \rho_{\psi'}\psi',
\end{eqnarray}
where $\dot \gamma=d\gamma/dt$ and the nonlinear terms are
($\rho_{\psi'}=\psi'\bar\psi'$, $\vec j_{\psi'}= 
\frac{\hbar}{m}{\rm Im\,}\big\{\bar \psi'\vec \nabla\psi'\big\}$) 
\begin{eqnarray}
R_1&=&R_1[\rho_{\psi'},\vec j_{\psi'}]=
\frac{m}{\hbar}\frac{\vec \nabla\cdot\vec
j_{\psi'}}{\rho_{\psi'}}\\
 R_2&=&R_2[\rho_{\psi'},\vec j_{\psi'}]=
\frac{\Delta\rho_{\psi'}}{\rho_{\psi'}}\\
R_4&=&R_4[\rho_{\psi'},\vec j_{\psi'}]=
\frac{m}{\hbar}\frac{\vec
j_{\psi'}\cdot \vec \nabla\rho_{\psi'}}{\rho_{\psi'}^2}\\
R_5&=&R_5[\rho_{\psi'},\vec j_{\psi'}]=
\frac{\vec\nabla\rho_{\psi'}
\cdot \vec \nabla\rho_{\psi'}}{\rho_{\psi'}^2}.
\end{eqnarray}
For $\dot\gamma=0$ the equation is 1-homogeneous in $\psi'$. For
$\dot\gamma\neq 0$ the last term is the famous
Bia{\l}ynicki-Birula--Mycielski  logarithmic
nonlinearity \cite{bbm}
which is widely believed to be ruled out experimentally. Let us
note that this
cannot be the case, since this would mean that the above nonlinear
equation is ruled out as well, but this is simply
the linear Schr\"odinger equation in a nonlinear disguise! 
Similarly, had all those impossibility proofs been correct in
their generality, the equations of this type would have to
generate unphysical effects. But the point is that they do not
generate {\it any\/} new effects, provided one uses a consistently
modified interpretation of the gauge transformed theory. 

NGT of density matrices can be used to refute another argument
of this variety. It is often stated that at the fundamental
level the density matrices have
to satisfy a linear equation since otherwise the convexity
principle would be violated. By this it is meant that having two
density matrices $\rho_1(x,y)$ and $\rho_2(x,y)$ their convex
combination $p_1\rho_1(x,y)+p_2\rho_2(x,y)$ should also be a
solution of the same equation and this implies linearity of the
Liouville-von Neumann equation. However, even assuming that this
argument is correct, the assumption that what we
observe experimentally is determined only by the diagonal
elements $\rho(x,x)$ implies that we should impose the convexity requirement
only on the diagonal. An example of a nonlinear dynamics that
possesses this kind of convexity is obtained by a NGT of a
linear Liouville-von Neumann equation. Taking $\rho'(x,y)=
N_{\lambda,\gamma_c}[\rho](x,y)$ we find 
that 
\begin{eqnarray}
[p_1\rho_1+p_2\rho_2]'(x,y)&\neq&
p_1\rho'_1(x,y)
+
p_2\rho'_2(x,y)
\end{eqnarray}
but
\begin{eqnarray}
[p_1\rho_1+p_2\rho_2]'(x,x)&=&
p_1\rho_1(x,x)
+
p_2\rho_2(x,x).
\end{eqnarray}
Even very simple NGT of $\rho$ (say, with $\lambda=1$) lead to
$\rho'$ that satisfy very complicated nonlinear equations
(involving, for $\lambda=1$, 12 nonlinear terms) which,
nevertheless, results in a theory fully equivalent to
liner quantum mechanics. 
To distinguish between the linear theory and 
nonlinear theories that satisfy the ``convexity principle on the diagonal"
one has to invent a measurement that is not reducible to a
measurement of particle's position.

\medskip
It is a real pleasure to acknowledge the discussions I had on the
subject with H.-D.~Doebner, J.~Hennig, W.~L\"ucke, and P.~Nattermann
when I was a DAAD fellow in Arnold Sommerfeld
Institute in Clausthal. I would like to especially thank G.~A.~Goldin who
explained me the physical background of nonlinear gauge
transformations and drawn my attention to the problem of
convexity. This little piece of work is supported by
the Polish-Flemish grant 007 and was completed during my stay in
CLEA, Brussels.

\end{document}